\documentclass{article}
\usepackage[T1]{fontenc}
\usepackage[utf8]{inputenc}
\usepackage{ismir} %
\usepackage{amsmath,cite,url}
\usepackage{graphicx}
\usepackage{color}

\title{IdolSongsJp Corpus: A Multi-Singer Song Corpus in the Style of Japanese Idol Groups}

\multauthor
  {Hitoshi Suda$^1$ \hspace{1cm} Junya Koguchi$^2$ \hspace{1cm} Shunsuke Yoshida$^3$}
  {{\bf Tomohiko Nakamura$^1$ \hspace{1cm} Satoru Fukayama$^1$ \hspace{1cm} Jun Ogata$^1$}\\
  $^1$ National Institute of Advanced Industrial Science and Technology (AIST), Tokyo, Japan\\
  $^2$ Meiji University, Tokyo, Japan \hspace{1cm} $^3$ The University of Tokyo, Tokyo, Japan\\
  {\tt\small suda.h@aist.go.jp, korguchi@gmail.com}
  }

\def\authorname{H. Suda, J. Koguchi, S. Yoshida, T. Nakamura, S. Fukayama, and J. Ogata}

\usepackage[bookmarks=false,pdfauthor={\authorname},pdfsubject={\pdfsubject},hidelinks]{hyperref}

\usepackage{amssymb,amsmath}
\usepackage{booktabs}
\usepackage[capitalise,noabbrev]{cleveref}
\usepackage{siunitx}

\makeatletter
\let\MYcaption\@makecaption
\makeatother
\usepackage{subcaption}
\makeatletter
\let\@makecaption\MYcaption
\makeatother

\begin{document}

\maketitle

\begin{abstract}

Japanese idol groups, comprising performers known as ``idols,'' are an indispensable part of Japanese pop culture. They frequently appear in live concerts and television programs, entertaining audiences with their singing and dancing. Similar to other J-pop songs, idol group music covers a wide range of styles, with various types of chord progressions and instrumental arrangements. These tracks often feature numerous instruments and employ complex mastering techniques, resulting in high signal loudness. Additionally, most songs include a song division (utawari) structure, in which members alternate between singing solos and performing together. Hence, these songs are well-suited for benchmarking various music information processing techniques such as singer diarization, music source separation, and automatic chord estimation under challenging conditions. Focusing on these characteristics, we constructed a song corpus titled IdolSongsJp by commissioning professional composers to create 15 tracks in the style of Japanese idol groups. This corpus includes not only mastered audio tracks but also stems for music source separation, dry vocal tracks, and chord annotations. This paper provides a detailed description of the corpus, demonstrates its diversity through comparisons with real-world idol group songs, and presents its application in evaluating several music information processing techniques.
\end{abstract}

\section{Introduction}

Processing music audio signals is one of the major directions in the music information retrieval (MIR) field\cite{Muller2011-nk,Purwins2019-vr}.
The techniques include beat tracking\cite{Goto2001-ex,Jean2003-om}, fundamental frequency estimation\cite{Goto2000-yv,Salamon2014-on}, automatic musical chord estimation\cite{Cheng2008-rz,Boulanger-Lewandowski2013-dh,Korzeniowski2016-lc}, music source separation (MSS)\cite{Virtanen2007-qh,Defossez2021-kw}, and automatic lyrics transcription (ALT)\cite{Mesaros2010-hf,Gao2022-uj}.

Training and evaluating MSS techniques in a supervised manner requires a corpus consisting of ground-truth stems, i.e., isolated acoustic signals categorized by instrument type.
Such corpora include MUSDB18\cite{Rafii2017-cs,Rafii2019-xr} and MoisesDB\cite{Pereira2023-ch}, which have been widely used for training and evaluating source separation methods\cite{Rouard2023-ee}.
On the other hand, some studies have shown that these corpora tend to exhibit lower loudness levels compared to commercially available tracks and have evaluated realistic performance on extended corpora by amplifying existing tracks to match commercial loudness levels\cite{Jeon2022-zr}.
However, some of the tracks in these corpora still feature fewer instruments and simpler mastering effects compared to contemporary commercial tracks, which makes practical performance evaluation challenging.
A more realistic assessment of MSS for contemporary songs requires constructing a music corpus that closely emulates the characteristics of these tracks.

In the field of music information processing, various techniques for multi-singer songs have also been explored, such as the separation of overlapping vocal signals\cite{Petermann2020-yw}.
Several corpora have been constructed for the study of such techniques, such as the jaCappella corpus and MedleyVox\cite{Nakamura2023-fp,Jeon2023-ot}.
In songs where singer combinations change between sections, key challenges include identifying which parts are sung by which singers and at what time.
In Japanese idol groups, such structures are referred to as song division (\textit{utawari} in Japanese), which is intentionally designed to enhance both the appeal of the song and the individuality of each idol\cite{Okada2013-rh,Suda2022-fy,Suda2024-dy}.
Such information is vital not only for music appreciation but also for applications such as music video production and live concert arrangements.
Similarly, for K-pop dance groups, \textit{line distribution} videos that visualize song division structures have been widely shared on platforms such as TikTok and YouTube.
The automatic recognition of these structures from music signals is known as singer diarization\cite{Thlithi2015-ds}.
Several studies have proposed specialized methods tailored for J-pop songs and have also constructed a dedicated corpus\cite{Suda2022-fy,Suda2024-dy}.
In summary, these studies suggest that focusing on multi-singer songs will further advance the field of music information processing.

Another important aspect of music information processing is active music listening, which involves interactive interfaces that enable listeners to better understand and appreciate music with the help of music understanding techniques\cite{Goto2007-ia}.
Studies on active music listening interfaces require preparing target songs and evaluating them through user interactions with the proposed interfaces.
While using commercial tracks for these evaluations is desirable, obtaining permission from individual copyright holders (e.g., record companies) is often challenging.
Similarly, some research-oriented song corpora, such as the RWC Music Database\cite{Goto2002-pv}, also require permission for public distribution, and conducting online evaluation experiments poses copyright concerns.
Thus, to advance research in this area, it is essential to construct a music corpus whose license explicitly permits research use and public sharing.

This study aims to construct a high-loudness corpus of multi-singer songs that can be distributed online and accessible to both researchers and ordinary listeners.
To address this challenge, we focus on Japanese idol group songs, which are well-suited for music information processing applications owing to their complex instrumental and vocal arrangements, commercial-level loudness, and diverse musical styles.
In addition, these idol group songs are indispensable elements of Japanese pop culture.
For instance, at the 2024 Japan Record Awards\footnote{\url{https://www.tbs.co.jp/recordaward/}}, FRUITS ZIPPER's ``NEW KAWAII'' received the Best Piece Award, and Cho Tokimeki$\heartsuit$Sendenbu's ``Saijokyu ni Kawaii no!'' received the Best Lyrics Award.
Given their cultural and technical importance, we commissioned professional creators to compose 15 songs in the style of Japanese idol groups and constructed a corpus titled IdolSongsJp.
These tracks are mastered to a loudness level comparable to commercial songs and feature realistic song division (\textit{utawari}) structures and diverse chord progressions.
Furthermore, the corpus includes not only mastered tracks but also stems designed for source separation, dry solo vocal tracks for each singer, and musical chord annotations, enabling a wide range of evaluations in music information processing.
To advance studies on music listening applications, we made this corpus publicly distributable for non-commercial use.
This paper presents the aim and detailed description of the corpus and demonstrates the diversity of the songs by comparing them with real-world idol group songs.
This paper also shows the application of the corpus by evaluating several fundamental MIR techniques.

\section{IdolSongsJp Corpus: A Multi-singer Corpus in the Japanese Idol Group Style}\label{sec:corpus}

We constructed a novel corpus titled IdolSongsJp, comprising 15 multi-singer songs in the style of Japanese idol groups.
This corpus features 10 female and 8 male singers, all of whom are professionals or semi-professionals, and each of the 15 songs features a unique combination of singers.
All creators are professionals with experience providing songs to real idol groups.
Each individual contributed to no more than three songs to ensure a diverse range of creative styles across the corpus.
Each song was designed to exhibit distinctive characteristics that reflect the diverse styles observed in Japanese idol group songs.
\Cref{tab:song_list} presents the list of songs along with their keywords and core concepts.
This corpus is available at \url{https://huggingface.co/datasets/imprt/idol-songs-jp}.

\begin{table*}
\centering
\setlength{\tabcolsep}{4pt}
\footnotesize
\begin{tabular}{llrccl}
\toprule
ID & Title & Tempo & No. of singers & No. of stems & Keywords and core concepts \\
\midrule
F01 & Introjuice & 186 bpm & 7 & 15 & Self-introduction; audience interaction (\textit{calls} and \textit{mixes}) \\
F02 & Aoharu Syndrome & 190 bpm & 6 & 11 & Rapidly unfolding lyrical narrative \\
F03 & Awake & 175 bpm & 5 & 12 & Frequent key changes (11 times) \\
F04 & Awanai Kagi & 123 bpm & 8 & 13 & Dance music with acoustic piano and guitar; questioning lyrics \\
F05 & Kaerimichi & 90 bpm & 6 & 14 & Heartbreak ballad; metaphorical lyrics; no multi-singer sections \\
F06 & Ima, Sekai wa Kagayaiteru & 132 bpm & 4 & 12 & Topics related to idol groups; metaphorical lyrics \\
F07 & Illuminyati & 160 bpm & 7 & 11 & Coined word; wordplay; audience interaction (chants) \\
F08 & tic-tac-toe & 120 bpm & 9 & 10 & Vocal doubling; vocoder effects; rap; code-switching with English \\
M01 & Liberty wing & 135 bpm & 8 & 15 & Introduction of other members; audience interaction (\textit{calls}) \\
M02 & trick star & 134 bpm & 7 & 17 & Augmented chords; high stem count \\
M03 & Answer & 185 bpm & 6 & 12 & Rock; ateji (alternate kanji pronunciations) \\ 
M04 & Uraomote Docchi? & 157 bpm & 7 & 10 & Release cut piano; Vocaloid style \\
M05 & Deep breath & 130 bpm & 5 & 8 & UK Garage; rap; code-switching with English; vocal chops \\
M06 & Soredemo, Sukidayo & 78 bpm & 4 & 11 & Heartbreak ballad; ad-libbed vocals and piano \\
M07 & ``Suki'' no Aizu & 138 bpm & 6 & 12 & Three members singing an octave lower; long spoken lines \\
\bottomrule
\end{tabular}
\caption{Songs in the IdolSongsJp corpus. The prefix of each song ID (F or M) indicates the gender of the singers.}
\label{tab:song_list}
\end{table*}

The songs exhibit several characteristics typical of Japanese idol songs, as described below:
\begin{itemize}
    \setlength{\parskip}{0pt}
    \item \textbf{Song division (\textit{utawari}) structures}. In this structure, the vocal arrangements vary across different sections of each song. Different singers take turns line by line, with some sections sung solo and others performed by multiple singers simultaneously. In addition, the songs feature countermelodies, harmonies, and choral sections (e.g., ``oohs'' and ``ahhs''), which are characteristic of J-pop songs.
    \item \textbf{Various music styles}. This corpus comprises a diverse range of music styles, including pop songs as well as genres such as UK Garage (M05), ballads (F05 and M06), rock (M03), and dance music (F08).
    \item \textbf{Various lyrical themes}. While love songs (e.g., F02, F04, M04, M06) are common in idol songs, some songs feature alternative themes, such as self-introduction (F01) and topics related to idol groups (F06).
    \item \textbf{High loudness}. The mastering process targeted a loudness of \SI{-7}{LUFS}, similar to that of commercial idol group songs. Since some creators release low-loudness versions for online platforms, such as Spotify and YouTube, this corpus also includes tracks mastered to a target loudness of \SI{-9}{LUFS}.
    \item \textbf{Reflection of Japanese idol culture}. At idol group concerts, audiences frequently cheer and shout in response to the performances. Specifically, tracks F01, F07, and M01 incorporate cheers, shouts, and chants (referred to as \textit{calls} and \textit{mixes} in Japanese\cite{Xie2021-hy}) into their recordings.
\end{itemize}

\begin{figure*}
    \centering
    \includegraphics[width=\linewidth,alt={Overview of the structure of the IdolSongsJp corpus}]{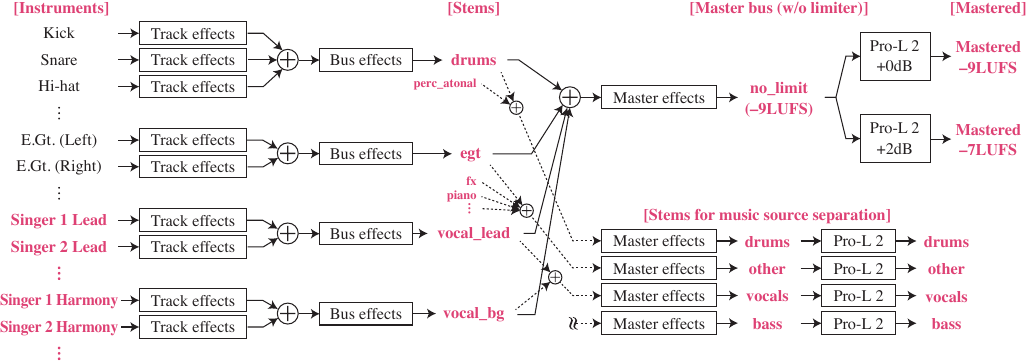}
    \caption{Overview of the production process for songs in the IdolSongsJp corpus. Data types included in the corpus are highlighted in red.}
    \label{fig:data_diagram}
\end{figure*}

This corpus comprises several types of data that are suitable for evaluating various music information processing techniques.
\Cref{fig:data_diagram} shows the data types organized according to the song production process.
All audio signals were sampled at \SI{48}{kHz} and encoded in 32-bit floating-point format to make dithering unnecessary.
The following list shows the included data:
\begin{itemize}
    \setlength{\parskip}{0pt}
    \item \textbf{Stems}. These stems were created by linearly mixing instrumental signals according to their respective categories and applying bus effects. The category definitions were derived from MoisesDB\cite{Pereira2023-ch}.
    \item \textbf{Stems for MSS}. This corpus employs four categories: drums, bass, vocals, and other. These signals were created by linearly mixing the corresponding stems and applying mastering effects.
    \item \textbf{Dry vocal tracks}.
    The corpus contains 414 individual dry vocal tracks, including lead melodies, countermelodies, harmonies, choruses, and chants.
    Each singer recorded the whole song, and final vocals in the mastered tracks were trimmed based on the designated song division structures.
    The tracks were processed using specialized software (e.g., Melodyne\footnote{\url{https://www.celemony.com/en/melodyne/what-is-melodyne}}) to adjust pitch and timing. These processed tracks can be utilized to train and evaluate singing voice synthesis techniques.
    \item \textbf{Master bus signals without limiters}. These signals were produced by linearly mixing the stems and applying mastering effects except the final limiter.
    \item \textbf{Mastered tracks} at \SI{-7}{LUFS} and \SI{-9}{LUFS}. For maximization and limiting, we used the Modern preset of FabFilter Pro-L 2\footnote{\url{https://www.fabfilter.com/products/pro-l-2-limiter-plug-in}}. The loudness of the tracks mastered to a target of \SI{-7}{LUFS} ranges between \SI{-7.1}{LUFS} and \SI{-7.0}{LUFS}, based on the ITU-R BS.1770-3 standard\cite{Radiocommunication-Sector-of-International-Telecommunication-Union-ITU-R-2012-ee}. The tracks have been designed and mastered based on the \SI{-7}{LUFS} version.
    \item \textbf{Off-vocal tracks and minus-one vocal tracks}. These signals were generated using the same process as for the mastered tracks, except that the vocal signals were excluded. Minus-one vocal tracks include backing harmonies and choruses. The corpus includes these versions in the same three mastering types as described above.
    \item \textbf{Solo versions}. Each of the 95 solo version tracks was mastered by mixing instrumental signals with the corresponding solo vocal signal, meaning that each track features the vocal performance of only one specific singer. Therefore, the corpus can also be utilized as a solo song corpus. Each track is available in the same three mastering types as described above.
    \item \textbf{Key and chord annotations} in Harte's shorthand notation\cite{Harte2005-xu} with time information.
    These annotations are based on the consensus of at least two hired expert annotators.
\end{itemize}

The authors retain all copyrights to the corpus.
This corpus is available free of charge for non-commercial research and entertainment purposes. No prior consent is required for such uses.
Any commercial use of the corpus requires prior permission from the authors.
Users may rearrange, parody, and apply machine learning techniques to the corpus, provided that the creators' moral rights are upheld.
To protect the rights associated with software instruments and commercial sample libraries, sampling the instrumental tracks to create unrelated content or to train machine learning models is prohibited.

\section{Comparison with Real-World Idol Group Songs}\label{sec:comparison}

The IdolSongsJp corpus is designed to capture the diverse styles of idol group music.
This section demonstrates the style diversity of the songs in our corpus by comparing their music embeddings to those derived from real-world idol group songs.

We processed 4,483 publicly available preview tracks performed by 234 female idol groups, which we obtained from a music subscription service.
To mitigate the effects of vocal characteristics and variations in the number of singers, we extracted the accompaniment signals using the fine-tuned model of Hybrid Transformer Demucs (HT Demucs)\cite{Defossez2021-kw,Rouard2023-ee}.
We then extracted 512-dimensional embeddings from these signals using a contrastive language--audio pretraining (CLAP) model\cite{Wu2022-nn}.
Subsequently, we reduced the dimensionality of these embeddings to two using uniform manifold approximation and projection (UMAP)\cite{McInnes2018-il}.
We also extracted embeddings from the songs in our corpus and reduced their dimensionality using the same UMAP parameters as previously described.
The CLAP model was \texttt{music\_audioset\_epoch\_15\_esc\_90.14}\footnote{\url{https://huggingface.co/lukewys/laion_clap}}, which is designed for music signals.
In this comparison, all real-world songs were performed by female groups, as these songs are generally more accessible and well-organized online than those by male groups.
Nonetheless, this section treats all songs in the IdolSongsJp corpus uniformly since vocal characteristics were mitigated.

\begin{figure}
    \centering
    \includegraphics[alt={A t-SNE visualization of music embeddings extracted from real-world idol group songs and the IdolSongsJp}]{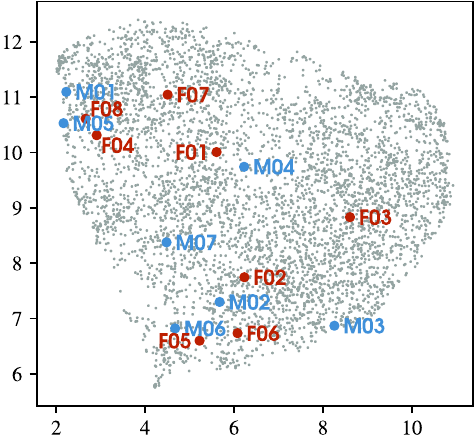}
    \caption{Visualization of music embeddings extracted from real-world idol group songs and songs from our corpus. Each gray dot represents a real-world idol group song, while red and blue dots indicate songs from our corpus featuring female and male vocals, respectively.}
    \label{fig:umap}
\end{figure}

\Cref{fig:umap} shows the results.
The embedding space reveals a continuous distribution of songs without forming distinct clusters.
The figure shows that the songs in our corpus are broadly distributed across the embedding space, indicating a wide variety of styles.
Tracks located near the periphery of the embedding space exhibit unique styles, while those near the center tend to have more common idol group characteristics.
This suggests that the corpus includes both distinctive songs (e.g., F05, F06, F08) and more typical ones (e.g., F01, M04).
Moreover, the absence of data points in certain regions of the embedding space, such as the upper right region, suggests directions for further expansion of the corpus.

\section{Application 1: Music Source Separation}

As mentioned in \cref{sec:corpus}, the IdolSongsJp corpus includes stem signals, which help the evaluation of MSS techniques.
In this section, we evaluate the performance of the fine-tuned model of HT Demucs\cite{Defossez2021-kw,Rouard2023-ee}, which separates music signals into four stems: bass, drums, vocals, and other.
We evaluated the performance using three types of input signals, all of which are provided in the corpus:
\begin{enumerate}
    \setlength{\parskip}{0pt}
    \item \textbf{Linear summation of stems without mastering effects}. The stems are summed linearly without any additional mastering processing.
    \item \textbf{Mastered tracks at $\mathbf{\SI[propagate-math-font=true]{-9}{LUFS}}$}. For these tracks, the same mastering effects as those used in producing the mastered tracks were applied to the ground-truth stems.
    \item \textbf{Mastered tracks at $\mathbf{\SI[propagate-math-font=true]{-7}{LUFS}}$}. The only difference from condition 2 is the gain parameter applied to the final limiter.
\end{enumerate}
Since mastering effects include not only maximizers and limiters but also equalizers, stereo imaging plug-ins, and other processing plug-ins, the final mixed signals differ considerably from the raw stems.
To address this discrepancy, we applied the same mastering effects to the individual stems in conditions 2 and 3 so that the acoustic conditions of the reference stems approximately matched those of the mastered tracks.
Note that since most mastering effects are nonlinear, simply summing the processed stems does not reproduce the final mastered tracks.
These effects were applied to each stem solely to approximate the overall acoustic conditions, such as loudness and stereo balances.
In this evaluation, signal-to-distortion ratio (SDR) was used as the evaluation metric.

\begin{figure}
    \centering
    \includegraphics[alt={SDR for separated signals from the IdolSongsJp corpus}]{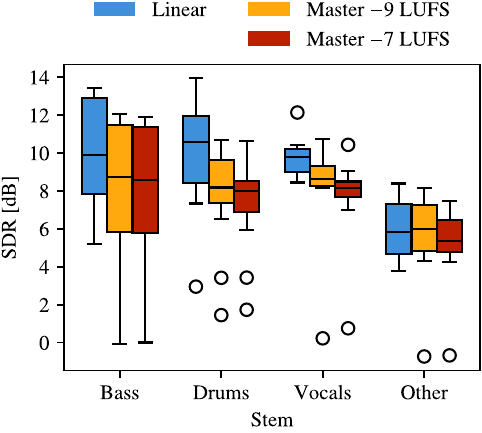}
    \caption{Signal-to-distortion ratios (SDR) for separated signals from the IdolSongsJp corpus, obtained using the fine-tuned model of HT Demucs\cite{Defossez2021-kw,Rouard2023-ee}. Dots indicate outliers that lie more than 1.5 times the interquartile range away from the quartiles.}
    \label{fig:separation_results}
\end{figure}

\Cref{fig:separation_results} shows the separation results.
For condition 1 (linear summation of stems), the separation performance was comparable to that observed with MUSDB18-HQ\cite{Rafii2019-xr}, demonstrating the effectiveness of the HT Demucs model.
In contrast, when processing the mastered signals, the separation performance deteriorated, particularly for drums and vocals, and was lower than previously reported for existing high-loudness music corpora\cite{Jeon2022-zr}.
One possible explanation is that certain mastering effects are designed to smooth, glue, and saturate the tracks and induce interactions between stems, thereby complicating MSS.
Notably, the mastered signals for track M05, ``Deep breath,'' exhibited the lowest separation performance across all stems.
Because this song is based on UK Garage, its sound design, especially the bass, differs considerably from the typical one found in pop and rock music.
These findings suggest that enhancing the diversity of genres and effects in the training tracks may be necessary to further improve the performance of MSS techniques.
Furthermore, the results indicate potential limitations in the current categorization of instruments, suggesting that target source separation techniques\cite{Liu2022-kz} will be required for more effective applications.

\section{Application 2: Automatic Chord Estimation}

\begin{figure}
    \centering
    \includegraphics[alt={Occurence rate of the chord qualities in the IdolSongsJp corpus}]{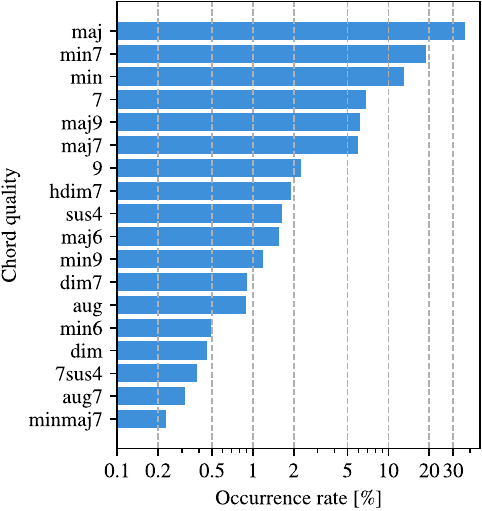}
    \caption{Distribution of chord qualities in the IdolSongsJp corpus, represented using Harte's shorthand notation\cite{Harte2005-xu}. Chord roots, inversions, tension notes, and non-chorded segments are excluded. The 7sus4 chord, which is not defined in this notation or in the open-source evaluation package \texttt{mir\_eval}, is composed of (1, 4, 5, $\flat$7).}
    \label{fig:chord_freq}
\end{figure}

As mentioned in \cref{sec:corpus}, the IdolSongsJp corpus contains musical chord annotations provided by expert annotators.
\Cref{fig:chord_freq} shows the occurrence rates of chord qualities, excluding chord roots, inversions, tensions, and non-chorded sections.
The proportion of major and minor chords is 50\%, which is significantly lower than the 65\% observed in the McGill Billboard corpus\cite{Burgoyne2011-rs}, a reference corpus used in the Music Information Retrieval Evaluation eXchange (MIREX) competition.
In contrast, the corpus includes a large number of tetrads and pentads (e.g., seventh and ninth chords), as well as diminished (dim) and augmented (aug) chords.
Therefore, effective chord estimation for this corpus requires a model with a large chord vocabulary capable of covering these diverse chord types.

\begin{figure}
    \centering
    \includegraphics[alt={Chord estimation accuracies}]{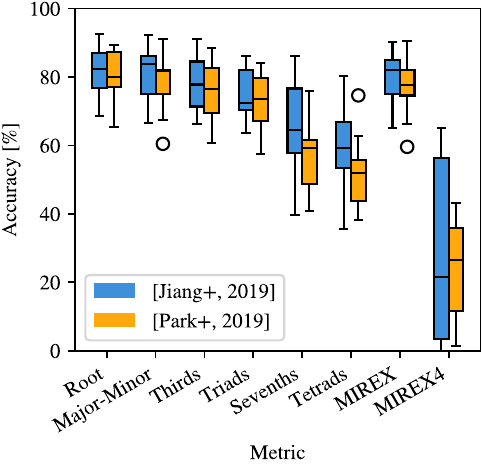}
    \caption{Chord estimation accuracies using two open-source methods\cite{Jiang2019-wr,Park2019-it}. Evaluation metrics are based on the Music Information Retrieval Evaluation eXchange (MIREX) competition and the implementation in the open-source package \texttt{mir\_eval}. The MIREX4 column indicates the recognition accuracy for chords consisting of four or more notes, where at least four component notes must be correctly identified. Dots indicate outliers that lie more than 1.5 times the interquartile range away from the quartiles.}
    \label{fig:chord_results}
\end{figure}

We applied and evaluated two open-source chord estimation methods on the IdolSongsJp corpus.
The first method is based on multitask classification of chord attributes (e.g., root, quality, bass, and seventh) using bi-directional long short-term memory (Bi-LSTM) networks\footnote{\url{https://github.com/music-x-lab/ISMIR2019-Large-Vocabulary-Chord-Recognition}}\cite{Jiang2019-wr}.
The second is an end-to-end method based on bi-directional Transformers\footnote{\url{https://github.com/jayg996/BTC-ISMIR19}}\cite{Park2019-it}.
Both methods can handle recognition tasks involving a large chord vocabulary.
\Cref{fig:chord_results} shows the chord estimation results.
For root note estimation and major/minor chord estimation, both methods achieved accuracies exceeding 80\%, demonstrating their effectiveness on this corpus.
However, the recognition accuracy for tetrads was below 60\% for both methods, indicating limitations in handling larger chord vocabulary.
In the figure, the MIREX4 column represents the recognition accuracy for chords consisting of four or more notes, where at least four component notes must be correctly identified.
In this case, the accuracy was below 30\% for both methods, indicating a large room for improvement in recognizing chords with complex structures.
In summary, chord estimation involving an extended chord vocabulary, such as tetrads and beyond, remains challenging for current methods.

\section{Application 3: Automatic Lyrics Transcription}

As described in \cref{sec:corpus,sec:comparison}, the songs in the IdolSongsJp corpus cover a wide range of musical styles with a variety of genres, tempos, and lyrical themes.
Therefore, the corpus serves as a benchmark dataset for various natural language processing tasks involving lyrics in diverse musical contexts.
For example, this section evaluates the performance of automatic lyrics transcription (ALT) using existing automatic speech recognition (ASR) techniques.

We utilized two ASR methods to perform ALT.
The first method uses Whisper with its large model\cite{Radford2022-jo}.
The language tag for Japanese was provided as input to its Transformer decoder.
The second method is based on a Conformer model\cite{Gulati2020-sg} that accepts Hidden-Unit BERT (HuBERT)\cite{Hsu2021-mm} features.
For the HuBERT model, we used \texttt{kushinada-hubert-large}\footnote{\url{https://huggingface.co/imprt/kushinada-hubert-large}}\cite{Takizawa2025-yo}.
The Conformer model was trained on LaboroTVSpeech, a large-scale Japanese speech corpus derived from television broadcasts\cite{Ando2021-lo}.
Both methods were designed for general ASR and were not specifically tailored for ALT.
In this evaluation, we used three types of input signals: (1) mastered tracks at \SI{-7}{LUFS}, (2) vocal signals separated from the mastered tracks using the fine-tuned model of HT Demucs, and (3) raw lead vocal stems without any mastering effects.

\begin{table}
    \centering
    \setlength{\tabcolsep}{5pt}
    \begin{tabular}{lrrrr}
        \toprule
        & $S$ [\%] & $D$ [\%] & $I$ [\%] & CER [\%] \\
        \midrule
        \multicolumn{5}{l}{Whisper\cite{Radford2022-jo}} \\
        \quad Mastered tracks & 10.3 & 8.7 & 21.8 & 40.7 \\
        \qquad +Demucs & 8.1 & 15.7 & 20.0 & 43.8 \\
        \quad Lead vocals & 7.9 & 14.1 & 17.7 & 39.7 \\
        \midrule
        \multicolumn{5}{l}{HuBERT+Conformer\cite{Gulati2020-sg,Hsu2021-mm,Takizawa2025-yo}} \\
        \quad Mastered tracks & 13.4 & 59.8 & 2.5 & 75.6 \\
        \qquad +Demucs & 16.3 & 26.1 & 5.1 & 47.5 \\
        \quad Lead vocals & 15.1 & 21.5 & 5.2 & 41.8 \\
        \bottomrule
    \end{tabular}
    \caption{Character error rates (CERs) for automatic lyrics transcription. Columns $S$, $D$, and $I$ represent substitution, deletion, and insertion error rates, respectively. The CER column shows the total character error rate, the sum of these three error types. Rows labeled Mastered tracks correspond to tracks mastered at \SI{-7}{LUFS}; +Demucs indicates separated vocal signals obtained from these mastered tracks using HT Demucs; and Lead vocals indicates the raw lead vocal stems without mastering effects.}
    \label{tab:cer}
\end{table}

\Cref{tab:cer} presents the character error rates (CERs) for the transcription results.
The performance of the Whisper model remained consistent across all three input conditions.
This consistency, as reported also in previous studies\cite{Cifka2024-vs}, demonstrates the robustness of Whisper against accompaniment signals and degradation caused by MSS.
In contrast, the results indicate that the Conformer model with HuBERT features was more negatively affected by accompaniment signals, despite having been trained on data that included music TV programs.
The two systems demonstrated different error tendencies: the Whisper model tended to produce insertion errors, while the Conformer model frequently generated deletion errors, with insertion errors occurring rarely.
Moreover, the Whisper model often produced hallucinated outputs, such as ``thank you for your watching.''
Conversely, the Conformer model tended to skip over fast segments and failed to recognize English phrases accurately, which contributed to its frequent deletion errors.

\section{Conclusion}

We constructed the IdolSongsJp corpus, a novel multi-singer song corpus in the style of Japanese idol groups.
The corpus includes not only mastered tracks but also stems for evaluating MSS techniques, dry vocal tracks, solo versions for all song--singer pairs, and chord annotations provided by expert annotators.
The songs cover a wide range of musical styles observed in Japanese idol groups, encompassing diverse genres, tempos, song division patterns, and lyrical themes.
Therefore, the corpus serves as a realistic resource not only for general tasks, such as MSS and automatic chord estimation, but also for song-specific applications, such as audience interaction enhancement and piano arrangement generation.

The IdolSongsJp corpus explicitly focuses on song division (\textit{utawari}) structures, making it well-suited for multi-singer tasks such as singer diarization and multi-pitch detection.
Future directions include applying this corpus to the development and evaluation of such tasks specific to multi-singer songs.
Furthermore, extending this approach to popular music genres from other regions, such as Korean and Chinese pop, may further broaden the applicability of music information processing techniques, particularly those for multi-singer songs.

\section{Acknowledgments}
This research was partially supported by the AIST policy-based budget project ``R\&D on Generative AI Foundation Models for the Physical Domain.''
The authors would like to acknowledge Mr. Takizawa (AIST) for his support in the speech recognition evaluation.

\section{Ethics Statement}
This corpus includes instrument stems and dry vocal tracks, which may be used for music generation and related applications. 
Instrument stems may be utilized to create new musical content through music generation techniques or sampling, potentially infringing on the rights of these sources.
Vocal signals may also enable synthetic speech that is unethical or impersonates others.
To address these concerns, such uses are explicitly prohibited under the license provided at the distribution repository.

\bibliography{paperpile}

\end{document}